\documentclass[twocolumn,showpacs,preprintnumbers,amsmath,amssymb]{revtex4}

\usepackage{graphicx}
\usepackage{dcolumn}
\usepackage{bm}
\usepackage{amsmath}

\newcommand{\MeV}{\mathrm{MeV}}
\newcommand{\GeV}{\mathrm{GeV}}

\newcommand{\meter}{\mathrm{m}}
\newcommand{\fm}{\mathrm{fm}}
\newcommand{\cm}{\mathrm{cm}}
\newcommand{\km}{\mathrm{km}}
\newcommand{\gram}{\mathrm{g}}
\newcommand{\second}{\mathrm{s}}
\newcommand{\AU}{\mathrm{AU}}
\newcommand{\Mpc}{\mathrm{Mpc}}

\newcommand{\nuc}{\mathrm{nuc}}
\newcommand{\const}{\mathrm{const}}

\begin{document}


\title{Color-charged Quark Matter in Astrophysics?}

\author{Congxin Qiu$^1$, Renxin Xu$^2$}
\affiliation{$^1$School of Space and Earth Sciences, Peking
University, Beijing 100871; o\underline{ }x\underline{ }o@yeah.net\\
$^2$School of Physics, Peking University, Beijing 100871;
r.x.xu@pku.edu.cn}

\date{\today}

\begin{abstract}
Color confinement is only a supposition, which has not been proved
in QCD yet.
It is proposed here that macroscopic quark gluon plasma in
astrophysics could hardly maintain colorless because of causality.
The authors expected that the existence of chromatic strange quark
stars as well as chromatic strangelets preserved from the QCD
phase transition in the early universe could be unavoidable if
their colorless correspondents do exist.
\end{abstract}

\pacs{12.38.Aw, 12.38.Mh, 97.60.Jd, 98.80.Cq}

\maketitle

The elementary strong interaction is believed to be recognized by
two distinct features: asymptotic freedom and color confinement.
Though Politzer~\cite{Politzer} and Gross and Wilczek~\cite{Gross}
proved asymptotic freedom in Quantum Chromodynamics (QCD), QCD in
nonperturbative regime is still unsolved and becomes one of the
top challenges for physics today.
Nambu~\cite{Nambu} discussed the possibility of color confinement
since the QCD vacuum is supposed to be a condensation of virtual
quark--antiquark pairs and gluons. In the strong coupling
approximation, lattice gauge calculation shows that the QCD
potential is linear in the infrared region~\cite{Cheng_GT}.
Certainly, we have never found a chromatic particle in an
accelerator experiment. However, unfortunately, all that evidences
mentioned above should not be enough to convince us a strictly
hold nature of color confinement, since infrared and ultraviolet
regions could be separated by one or more discontinuous phase
transitions~\cite{Cheng_GT}.

\begin{figure}
\resizebox{7cm}{7cm}{\includegraphics{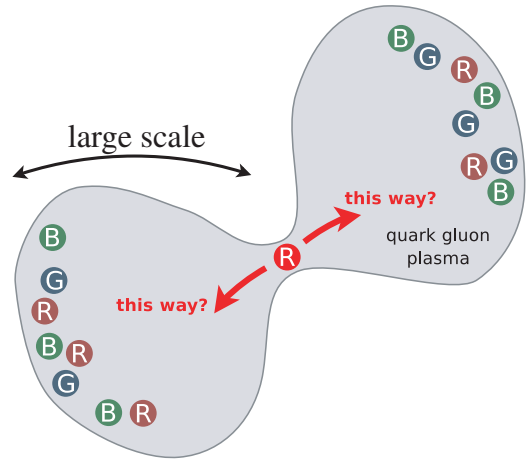}}
\caption{\label{fig:split}An illustration of splitting a large
bulk of quark matter. If a red quark in the center (labelled as
``{\bf R}'') wants to decide which way to go, it should be able to
count {\em instantaneously} the color-charges of quarks in at
least one side in order to keep the divided two parts in
color-singlet states. It could then be very difficult to maintain
color-singlets of the splitting pieces in a macroscopic scale.}
\end{figure}

The special QCD vacuum may probably explain the color-singlet of
particles in the high energy experiments.
Color-charged quarks, exchanging madly gluons, are supposed to be
confined in color-neutral hadrons (baryons or mesons) with others.
When one of the quarks in a given particle is ``pulled'' away from
its neighbors, it would be energetically favorable for the virtual
quark--antiquark pairs to become valency and to keep the hadrons
in singlet states again.
However, things would be much different in case of a huge bulk of
astrophysical quark gluon plasma.
Let's consider a bulk of quark matter which is initially
colorless. It should evaporate or split if it has to reduce its
volume for certain reasons. In case of evaporating particles
(e.g., baryons or mesons), color-singlet might keep easily since
this evaporation could be considered as a {\em local}
process \footnote{%
Note that QCD is a local gauge theory, and color confinement would
then be a local concept.
}. But in the splitting process, it could be hard for us to
believe that every splitting pieces {\em happen} to be colorless
because, in this non-local case, the time needed for transforming
information (even as fast as light) from one part of the bulk to
the other should not be negligible (Fig.~\ref{fig:split}).

There is a great difference between electroneutral and colorless
states.
An electrically charged body may discharge via ejecting electrons
or operating virtual $e^\pm$ pairs in the vacuum. How about
chromatic quark matter?
Note that the QED vacuum is screening, while the QCD vacuum is
anti-screening. It might be energetically favorable for chromatic
quark matter to locally excite and eject color-singlet particles,
and the color-charge keeps then.
The interaction between chromatic bulks of quark matter might only
be in very short distance {\em if} single color-gluon interchange
is not allowed since mesons or glueballs as intermediate particles
are very massive. All the observations above could show us that it
would be hard for chromatic quark matter to drop its color off
unless by random collision.

If color confinement is not \emph{exactly} hold, chromatic quark
matter could be in an extremely high energy scale, which should be
too high for nowadays accelerators to achieve.
Surely, the energy scale would be hard to estimate in the
framework of quantum field theory since the matter is in a
non--local and bound state.
We have then tried two phenomenological models~\cite{Qiu} to make
sense of this. In the first model, we considered a bulb in a color
dia--electric medium~\cite{TDLee_PH} which has color dielectric
constant less than one. The bulb can change its radius to minimize
its total energy. And in the second model, we simply laid up a lot
of color dipoles in the vacuum and calculate their respondence for
a valance color charge in the center.
Both the models show us that the minimum energy scale, $U_{\min}$,
for exciting chromatic quark matter should depend on the radius,
$R$, of the bulk, with a likely relation of $U_{\min} \simeq
\const \cdot 1 / R$. The constant in this relation can be
estimated if we pay attention to the fact that no accelerator
(which have a energy scale of about $100 \GeV$) is powerful enough
to create a chromatic particle (which have a typical radius of
about $1 \fm$). The constant could then be greater than $\sim
10^{-10} \MeV \cdot \meter$.
It is observed then that color confinement can only be hold
exactly only if the constant is infinity. Therefore, the
conclusion could be: \emph{if color confinement is exactly hold
for microscopic particles ($\sim 1$ fm), it is also exactly hold
for a bulk of quark matter; but if it is just approximately hold,
to create a bulk of chromatic quark matter is much easier (i.e.,
to need lower energy) than to create a chromatic microscopic
particle.}

The first implication of chromatic quark matter could be of
splitting strange stars.
Strange stars are potential candidates for the nature of
pulsar-like stars~(\cite{Haensel,Alcock}, see \cite{xu06} for a
review). They are a kind of fermion stars, which are bound by
strong interaction (and gravity if stellar masses approach the
maximum limit) and are in fact large bulks of quark gluon plasma.
There could be two conceivable channels to split bulk quark
matter: merge of binary strange stars and supernova explosion. It
is very uncertain to calculate the former process, and we then
think about the later one only as following.

\begin{figure}
\resizebox{8cm}{6.4cm}{\includegraphics{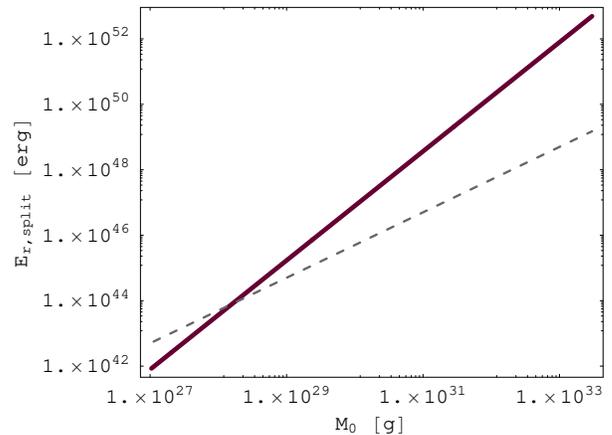}}
\caption{\label{pic:rSplit}Comparison of the rotation energy
needed for a star to split (solid line) and the rotation energy
the star may actually have (dashed line) for varying stellar mass
($M_0$).
In the calculation, we choose $\rho = 4.1 \times 10^{14} {\rm g}
\cdot \cm^{-3}$ for low mass strange stars. It is shown that a
strange quark star to be as low as $10^{-6}M_\odot$ could split.}
\end{figure}

Based on the kick velocity, $v_{\rm k}$, of pulsars, which we
choose $1000 \km \cdot \second^{-1}$ as an upper
limit~\cite{Hansen}, the kinematic energy scale of supernova
explosion can be estimated.
Assuming an equipartition rule for stellar rotational energy and
kick energy, we could then estimate the critical condition for
splitting a strange star~\cite{Qiu}.
The relation between spin frequency and equilibrium shape of a
liquid star was given by Maclaurin in the framework of Newtonian
gravity, with a equation~\cite{Chandrasekhar} of
\begin{equation}
    \Omega^{2} = 2 \pi G \rho \left[ \frac{\sqrt{1 - e^2}}{e^3} (3 - 2 e^2) \arcsin e
        - \frac{3 (1 - e^2)}{e^2} \right],
\end{equation}
where $\rho$ is the density of the star, $\Omega$ is the angular
velocity, $e$ is the eccentricity of the meridian plane, and $G$
is the Newtonian gravitational constant.
A rotating star would disintegrate at the bifurcation point of the
Jacobian sequence branches, with $e =
0.8127$~\cite{Chandrasekhar_Maclaurin,Chandrasekhar_Jacobi}.
Assuming $I\Omega^2=Mv_{\rm k}^2$, one can calculate the
rotational energies for different stellar masses, $M$, where $I$
is the momentum of inertia and $v_{\rm k}=10^8$ cm/s. In
Fig.~\ref{pic:rSplit}, the rotation energy ($E_{\rm r, split}$) of
a star with $e = 0.8127$ is shown by a solid line (i.e., $e$ is
fixed in Equ.(1)), whereas the actual rotation energy is shown by
a dashed line (i.e., $\Omega$ is determined by $v_{\rm k}$).
Comparing those two kinds of rotation energies, we could see that
a strange quark star with initial matter of $\sim 10^{-5} M_\odot$
may split into two or more bulks of quark matter.

The calculation shown in Fig.~\ref{pic:rSplit} can only applicable
to rotating rigid body kicked during in the supernova explosion,
but a real star should not be in this situation.
During the birth of protoquark stars with high temperature, $T$,
and high $\nabla T$~\cite{Woosley}, hot turbulently convective
quark stars may eject low-mass quark matter (quark
nuggets)~\cite{xu06b}.
Also during supernova explosions, small quark nuggets could
possibly be kicked out from the protoquark stars if they are not
rigid rotators. For the sake of simplicity, we assumed that the
energy kicked to the small piece is the same as to the quark star
(which is related to the observable kick velocity of pulsars). To
compensate the gravitational potential energy (for a quark star
not too massive, the typical density is a constant) in the
framework of Newtonian, the maximum piece can be kicked to
infinity has a typical mass $M = 7.5 \times 10^{28} \gram$ and
radius $R = 0.35 \km$ (for a constant density of $\rho = 4.1
\times 10^{14} \gram \cdot \cm^{-3}$).
In the calculation above we neglect the energy loss caused by
gravitational radiation, which is very effective~\cite{Qiu} but
hard to calculate. The maximum mass of splitting pieces might then
be actually smaller than $\sim 10^{28}$ g. This kind of splitting
pieces of quark matter could be candidates for the planets of
pulsars~\cite{Wolszczan}.
However, it is worth noting that, even though the kicked nuggets
take color-charges, the colors might not have dynamical contrition
to be as strong as gravity if chromatic quark matter would not
interact in a long distant.

The second implication could be about the chromatic strangelets
preserved during the QCD phase transition of the early Universe.
In the standard model~\cite{Boyanovsky}, a first-order QCD
transition leads to bubble nucleation. Bubbles of quark--gluon
plasma form during the transition.
In case of homogeneous nucleation, the mean distance of
nucleation, $d_{\nuc}$, could be $<2 \cm$, while the value of
$d_{\nuc}$ may be several meters in the case of heterogeneous
nucleation~\cite{Christiansen}.
The strangelets proposed by Witten~\cite{Witten} seems unlikely
since that suggestion needs $d_{\nuc} \gtrsim 300
\meter$~\cite{Boyanovsky}. However, things could be much different
if chromatic strangelets were created during cosmic QCD
phase-transition.
As we have mentioned, it could be energetically favorable to exist
huge bulk of chromatic quark matter since $U_{\rm min} \propto
R^{-1}$. In this sense, the hadronization of chromatic quark
nuggets could not be very effective and may still be residual
today.

Let's estimate roughly the number density of such strangelets in
the Universe, assuming that all the strangelets take color-charges
and could survived.
Lattice gauge calculations shows us that the QCD phase transition
occurs at a temperature of $T_{\mathrm{c}} = 170
\MeV$~\cite{Boyanovsky}. Using the relation of ``$a(t) \cdot T(t)
= \const$'' for radiation field in the expanding Universe, where
the scale factor $a(t)$ and the temperature $T(t)$ are function of
cosmic time $t$, and choosing $d_{\nuc} = 2 \cm$ and $T({\rm
today}) = 2.73$ K of cosmic microwave background, we obtain a
number density of $\sim (0.1 \AU)^{-3}$ in today's Universe.

It is not easy to estimate the mass of that kind of strangelets
since one can not have a believable method to calculate the total
energy of chromatic quark matter. A possible restriction for the
mass could be obtained by noting that the total mass of these
strangelets should be lower than that of dark matter.
If we choose the Hubble constant $H_{0} = 72 \km \cdot
\second^{-1} \cdot \Mpc^{-1}$ and assume that $25\%$ of the total
energy density of the Universe is in the form of dark matter in
the concordance model~\cite{Boyanovsky}, we could have an upper
limit of the strangelets' mass: $m_{\rm s} \lesssim 7.4 \times
10^{6} \gram$.
The radius of strangelets with mass of $\sim 10^{6} \gram$ is
about $10^{-3} \cm$. It is possible that such strangelets, which
are lighter than asteroids, are wandering in our solar system, but
we can hardly observe them since they seems to contribute only
gravitational interaction in the space.
Nevertheless, these strangelet bummers may have had significant
consequence during the early Universe (e.g., the period of galaxy
formation).

Is there any experimental feature of chromatic quark matter? This
is really an interesting question to be answered by future more
investigations.
Similar to the Schwinger process~\cite{Schwinger} of the QED
vacuum polarization by a prescribed electromagnetic field, a
highly color-charged quark matter may radiate colorless hadrons in
the nearby polarized QCD vacuum, but could become more and more
difficult and may finally stop when its mass decreases to a
critical value due to the fact of $U_{\rm min}\propto 1/R$.
A particularly fascinating process could be that a chromatic
relativistic strangelet goes into the Earth's atmosphere. It may
absorb nucleons at first, and then its temperature increases to be
high enough to evaporate hadrons. The character of its atmospheric
shower depends on the detail interactions, which is certainly
model-dependent. Are some cosmic events (e..g., the ultra-high
energy cosmic rays) related to chromatic strangelets? We can not
know at this time.

Let's summarize briefly our opinion. Chromatic strange quark stars
as well as strangelets could be unavoidable if their colorless
correspondents do exist, but evidence for color-charges might
hardly be obtained nowadays. We suggest that color confinement
would not be hold exactly in the nature.

This work is supported by NSFC (10573002) and the Key Grant
Project of Chinese Ministry of Education (305001). It is also part
of the thesis by one of the authors (Congxin Qiu).

\end{document}